\documentstyle[12pt,epsfig]{article}

\newlength{\dinwidth}
\newlength{\dinmargin}
\def\lapproxeq{\lower .7ex\hbox{$\;\stackrel{\textstyle
<}{\sim}\;$}}
\def\gapproxeq{\lower .7ex\hbox{$\;\stackrel{\textstyle
>}{\sim}\;$}}
\def\gtrsim{ \;\raisebox{-.7ex}{$\stackrel{\textstyle
>}{\sim}$}\; }
\def\lesim{ \;\raisebox{-.7ex}{$\stackrel{\textstyle
<}{\sim}$}\; }

\def\be{\begin{equation}}
\def\ee{\end{equation}}
\def\bea{\begin{eqnarray}}
\def\eea{\end{eqnarray}}
\newcommand{\bel}[1]{\begin{equation}\label{#1}}
\newcommand{\beal}[1]{\begin{eqnarray}\label{#1}}

\def\bb{b\bar{b}}
\def\cc{c\bar{c}}
\def\qq{q\bar{q}}
\def\pp{p\bar{p}}
\def\ra{ \rightarrow }

\def\GeV{{\rm GeV}}

\begin{document}
\begin{flushright}
IPPP/03/10 \\
DCPT/03/20 \\
13 March 2003 \\
\end{flushright}

\vspace*{2cm}

\begin{center}
{\Large \bf Diffraction of Protons and Nuclei at High
\renewcommand{\thefootnote}{\fnsymbol{footnote}}Energies\footnote[2]{To
appear in the special issue of Acta Physica Polonica to celebrate
the 65th Birthday of Professor Jan
Kwieci\'nski.}\renewcommand{\thefootnote}{\arabic{footnote}}\setcounter{footnote}{0}
}

\vspace*{1cm}
\textsc{A.B.~Kaidalov$^{a,b}$, V.A.~Khoze$^{a,c}$, A.D. Martin$^a$ and M.G. Ryskin$^{a,c}$} \\

\vspace*{0.5cm} $^a$ Department of Physics and Institute for
Particle Physics Phenomenology, \\
University of Durham, DH1 3LE, UK \\
$^b$ Institute of Theoretical and Experimental Physics, Moscow, 117259, Russia\\
$^c$ Petersburg Nuclear Physics Institute, Gatchina,
St.~Petersburg, 188300, Russia \\

\end{center}

\vspace*{1cm}

\begin{abstract}
We review the description of high energy diffraction from both the
$s$ and $t$ channel viewpoints, and demonstrate their consistency.
We emphasize the role played by $s$ channel unitarity and
multi-Pomeron exchanges. We explain how these effects suppress
hard diffractive processes. As examples, we describe the
calculation of diffractive dijet production at the Tevatron and
predict some of the rich diffractive phenomena accessible in
proton--nuclear collisions at RHIC.
\end{abstract}

\vspace*{1cm}


\section{Introduction}

A broad class of processes at high energies has properties
analogous to the classical pattern of the diffraction of light.
These are usually called diffractive processes. The classic
example is the elastic scattering of hadrons on nuclei, which has
an angular distribution with a series of minima and maxima,
analogous to the diffraction of light on a black disc. In
diffractive processes the wave nature of particles is clearly
revealed and the necessity of a quantum mechanical description in
terms of amplitudes, and not probabilities, is evident.

Unitarity relates the imaginary parts of forward elastic
scattering amplitudes to total cross sections. Hence studies of
elastic scattering and $\sigma_{\rm tot}$ for different targets
and projectiles was traditionally one of important source of
information on the strengths of interactions and their radii. In
particular, data on the high-energy interactions of hadrons
indicate that the radius of the strong interaction increases with
energy, and indicate that the total cross sections will continue
to increase in the energy region not explored at present
accelerators.

Processes of diffractive dissociation of colliding
particles~\cite{Feinberg} provide new possibilities for the
investigation of the dynamics of high-energy hadronic
interactions. These processes are dominated by soft,
nonpertubative aspects of QCD, and hence provide important
information on the dynamics. Indeed, they provide a rich testing
ground for models of soft interactions. In this paper we
demonstrate that the cross sections of inelastic diffractive
processes are very sensitive to nonlinear, unitarity effects.

A new class of diffractive phenomena---hard diffraction---has been
extensively studied both experimentally and theoretically in
recent years. These processes include the total cross section and
diffractive dissociation of a virtual photon at high energy at
HERA, and the diffractive production of jets, $W,Z$~bosons, heavy
quarks, etc. in hadronic collisions. The latter processes become
possible at very high (c.m.) energies ($\sqrt{s}$) because {\em
large} mass $M$ diffractive states may be produced with $M^2\ll
s$. Investigation of these processes gives new possibilities for
the study of the interplay of soft and hard dynamics in QCD. An
important question---the breaking of QCD factorization for hard
reactions in diffractive processes---will be discussed below.

Nowadays diffractive hard processes are attracting attention as a
way of extending the physics programme at proton colliders,
including novel ways of searching for New Physics, see, for
example, \cite{AR}--\cite{BPR03} and references therein. An
especially interesting process is the exclusive double diffractive
production of a Higgs boson at the LHC; $pp\ra p + H + p$, where
the $+$ signs denote the presence of large rapidity gaps. Clearly
a careful treatment of both the {\em soft} and {\em hard} QCD
effects is crucial for the reliability of the theoretical
predictions for these diffractive processes.

The outline of the article is as follows. In
Section~\ref{sec:theor} we review diffraction phenomena from both
the $s$ and $t$ channel viewpoints. In Section~2.3 we link these
approaches together. We emphasize the convenience of working in
terms of diffractive eigenstates, as originally proposed by Good
and Walker~\cite{Good}. In Section~3 we discuss how these
eigenstates may be identified with the different parton
configurations of the proton. Scattering through each of the
eigenstates is suppressed (by unitarity or multi-Pomeron effects)
by different amounts. In Section~4 we study some examples of hard
diffractive processes. First we summarize a calculation of
diffractive dijet production at the Tevatron, which shows that the
estimate of the overall suppression of the cross section is in
agreement with the data. In Section~4.2 we turn to diffractive
proton--nuclear processes. We present predictions for a range of
hard diffractive processes which are accessible at RHIC. Finally,
Section~5 contains some brief conclusions.

Since this article is an attempt at a short review\footnote{Other
reviews of diffraction can be found, for example,
in~\cite{Kaidiff}--\cite{DeWolf}.} of high energy diffractive
phenomena, we have only provided a few references to help the
reader. Hence our reference list is incomplete and we apologize to
those authors whose work has not been cited. Section~4 contains
new results for diffractive proton--nuclear processes.

\section{Theoretical approaches to diffraction at high
energies}\label{sec:theor}

The investigation of the diffractive scattering of hadrons gives
important information on the structure of hadrons and on their
interaction mechanisms. Diffractive processes may be studied from
either an $s$ channel or $t$ channel viewpoint.  We will discuss
these approaches in turn, and then show that they are
complementary to, and consistent with, each other.

\subsection{Diffraction from an $s$ channel viewpoint}

Unitarity plays a pivotal role in diffractive processes. The total
cross section is intimately related to the elastic scattering
amplitude and the scattering into inelastic final states via
$s$~channel unitarity, $SS^\dag = I$, or
\be {\rm disc}\ T \equiv T-T^\dag = iT^\dag T \label{eq:discT}\ee
with $S=I+iT$. If we were to focus, for example, on elastic
unitarity, then disc would simply denote the discontinuity of $T$
across the two-particle $s$ channel cut. At high energies,
$s$~channel unitarity relation is diagonal in the impact
parameter, $b$, basis, such that
\be 2 {\rm Im}\,T_{\rm el}(s,b) = |T_{\rm el}(s,b)|^2 + G_{\rm
inel}(s,b) \label{eq:a1} \ee
with
\bea \sigma_{\rm tot} & = & 2\int d^2b\, {\rm Im}\,T_{\rm el}(s,b)
\\
\sigma_{\rm el} & = & \int d^2b\,|T_{\rm el}(s,b)|^2 \\
\sigma_{\rm inel} & = & \int d^2b\,\left[2{\rm Im}\,T_{\rm
el}(s,b) - |T_{\rm el}(s,b)|^2\right]. \eea
The general solution of (\ref{eq:a1}) is
\bea T_{\rm el} & = & i(1-\eta e^{2i\delta}) \label{eq:elastamp}\\
G_{\rm inel} & = & 1-\eta^2, \label{eq:gensol} \eea
where $\eta(s,b)^2 = \exp(-\Omega(s,b))$ is the probability that
no inelastic scattering occurs. $\Omega\geq 0$ is called the
opacity (optical density) or eikonal\footnote{Sometimes $\Omega/2$
is called the eikonal.}.

The well known example of scattering by a black disc, with
$\eta=0$ for $b<R$, gives $\sigma_{\rm el} = \sigma_{\rm inel}
=\pi R^2$ and $\sigma_{\rm tot} = 2\pi R^2$. In general, we see
that the absorption of the initial wave due to the existence of
many inelastic channels leads, via $s$~channel unitarity, to
diffractive elastic scattering.

So much for elastic diffraction. Now we turn to inelastic
diffraction, which is a consequence of the {\em internal
structure} of hadrons. This is simplest to describe at high
energies, where the lifetimes of the hadronic fluctuations are
large, $\tau\sim E/m^2$, and during these time intervals the
corresponding Fock states can be considered as `frozen'. Each
hadronic constituent can undergo scattering and thus destroy the
coherence of the fluctuations. As a consequence, the outgoing
superposition of states will be different from the incident
particle, and will most likely contain multiparticle states, so we
will have {\em inelastic}, as well as elastic, diffraction.

To discuss inelastic diffraction, it is convenient to follow Good
and Walker~\cite{Good}, and to introduce states $\phi_k$ which
diagonalize the $T$ matrix. Such eigenstates only undergo elastic
scattering. Since there are no off-diagonal transitions
\be \langle \phi_j|T|\phi_k\rangle = 0\qquad{\rm for}\ j\neq k \ee
a state $k$ cannot diffractively dissociate in a state $j$. We
have noted that this is not true for hadronic states due to their
internal structure. One way of proceeding is to enlarge the set of
intermediate states, from just the single elastic channel, and to
introduce a multichannel eikonal. We will consider such an example
below, but first let us express the cross section in terms of the
probabilities $F_k$ of the hadronic process proceeding via the
various diffractive eigenstates $\phi_k$.

Let us denote the orthogonal matrix which diagonalizes ${\rm
Im}\,T$ by $a$, so that
\be \label{eq:a3} {\rm Im}\,T \; = \; aFa^T \quad\quad {\rm with}
\quad\quad \langle \phi_j |F| \phi_k \rangle \; = \; F_k \:
\delta_{jk}. \ee
Now consider the diffractive dissociation of an arbitrary incoming
state
\be \label{eq:a4} | i \rangle \; = \; \sum_k \: a_{ik} \: | \phi_k
\rangle. \ee
The elastic scattering amplitude for this state satisfies
\be \label{eq:a5} \langle i |{\rm Im}~T| i \rangle \; = \; \sum_k
\: |a_{ik}|^2 \: F_k \; = \; \langle F \rangle, \ee
where $F_k \equiv \langle \phi_k |F| \phi_k \rangle$ and where the
brackets of $\langle F \rangle$ mean that we take the average of
$F$ over the initial probability distribution of diffractive
eigenstates. After the diffractive scattering described by
$T_{fi}$, the final state $| f \rangle$ will, in general, be a
different superposition of eigenstates from that of $| i \rangle$,
which was shown in~(\ref{eq:a4}). At high energies we may neglect
the real parts of the diffractive amplitudes. Then, for cross
sections at a given impact parameter $b$, we have
\bea \label{eq:a6} \frac{d \sigma_{\rm tot}}{d^2 b} & = & 2 \:
{\rm Im} \langle i |T| i \rangle \; = \; 2 \: \sum_k
\: |a_{ik}|^2 \: F_k \; = \; 2 \langle F \rangle \nonumber \\
& & \nonumber \\
\frac{d \sigma_{\rm el}}{d^2 b} & = & \left | \langle i |T| i
\rangle \right |^2 \; = \; \left (
\sum_k \: |a_{ik}|^2 \: F_k \right )^2 \; = \; \langle F \rangle^2 \\
& & \nonumber \\
\frac{d \sigma_{\rm el \: + \: SD}}{d^2 b} & = & \sum_k \: \left |
\langle \phi_k |T| i \rangle \right |^2 \; = \; \sum_k \:
|a_{ik}|^2 \: F_k^2 \; = \; \langle F^2 \rangle. \nonumber \eea
It follows that the cross section for the single diffractive
dissociation of a proton,
\be \label{eq:a7} \frac{d \sigma_{\rm SD}}{d^2 b} \; = \; \langle
F^2 \rangle \: - \: \langle F \rangle^2, \ee
is given by the statistical dispersion in the absorption
probabilities of the diffractive eigenstates. Here the average is
taken over the components $k$ of the incoming proton which
dissociates. If the averages are taken over the components of both
of the incoming particles, then (\ref{eq:a7}) is the sum of the
cross section for single and double dissociation.

Note that if all the components $\phi_k$ of the incoming
diffractive state $| i \rangle$ were absorbed equally then the
diffracted superposition would be proportional to the incident one
and  the inelastic diffraction would be zero.  Thus if, at very
high energies, the amplitudes $F_k$ at small impact parameters are
equal to the black disk limit, $F_k = 1$, then diffractive
production will be equal to zero in this impact parameter domain
and so will only occur in the peripheral $b$ region. Such
behaviour already takes place in $pp$ (and $p\bar{p}$)
interactions at Tevatron energies. Hence the impact parameter
structure of inelastic and elastic diffraction is drastically
different in the presence of strong $s$ channel unitarity effects.

Under the assumption that  amplitudes $F_k$ at high energies can
not exceed the black disk limit, $F_k\leq 1$,
equations~(\ref{eq:a6}) lead to the following bound
\be \label{eq:PB} \frac{d \sigma_{\rm el \: + \: SD}}{d^2 b} \leq
\frac{1}{2} \frac{d \sigma_{\rm tot}}{d^2 b}\,, \ee
known as the Pumplin bound~\cite{Pumbound}.

A simple realistic application of the above framework is the
interaction of a highly virtual photon with a nucleus. For recent
discussions and references see, for example,~\cite{F,N}. The wave
function for the photon to fluctuate into a $\qq$ pair may be
written $\psi(\vec r,\alpha)$, where $\vec r$ is the separation of
the $q$ and $\bar q$ in the transverse plane and $\alpha$,
$1-\alpha$ are the longitudinal momentum fractions of the photon
momentum carried by the $q$ and $\bar q$. For small $r$ the cross
section for the $\qq$ pair to interact with a nucleon behaves as
$\sigma\sim\alpha_S r^2$. Thus, at fixed impact parameter $b$, the
opacity
\be \Omega\ =\ \sigma T_A(\vec b)\ \propto\ r^2 \label{eq:opacity}
\ee
where
\be T_A(b)\ =\ \int_{-\infty}^\infty dz\,\rho(\vec b, z)
\label{eq:opticaldensity} \ee
and $\rho(\vec b,z)$ is the density of nucleons in the nucleus
($\int T_A(b)\,d^2b = A$). $T_A(b)$ is often called the nuclear
density per unit area or the optical thickness of the nucleus.

At high energy, due to Lorentz time dilation, the lifetime of each
component with fixed $\vec r$ is much longer than the interaction
time of the $\qq$ pair inside the nucleus $A$. Hence the
$\gamma^*A$ amplitude is of the form\footnote{This result was
first proposed within QCD in Refs.~\cite{BERTSCH} and
\cite{coltransp}, but, long before, an analogous expression was
originally given~\cite{Feinberg} for deuteron scattering with
proton and neutron constituents in the place of the $\qq$
fluctuation of the photon.}
\be\langle 1-e^{-\Omega/2}\rangle\ =\ \int d^2r\,
d\alpha\,\psi^*(\vec r,\alpha)\left[1-\exp(-\sigma(\vec
r)T_A(b)/2)\right]\,\psi(\vec r,\alpha). \label{eq:gammastarA} \ee
On the other hand, at low energy, where the lifetime of the $\qq$
fluctuations is small, we have to average $\sigma(\vec r)$ already
in the exponent, since each nucleon interacts with a different
$\gamma^*\ra\qq$ fluctuation. Therefore the low energy amplitude
is
\be 1 - \exp(-\langle\Omega\rangle/2)\ =\ 1 -
\exp(-\langle\sigma\rangle T_A/2). \label{eq:lowenergyamplitude}
\ee
To simplify the discussion we, here, neglect the real part of the
elastic amplitude; that is we assume in (\ref{eq:elastamp}) that
$\delta\ll 1$. Modulo this assumption,
(\ref{eq:lowenergyamplitude}) is exactly the Glauber formula,
which was originally derived for an incoming particle of energy
$E\lesim1$~GeV, which therefore has enough time to return to its
equilibrium state between interactions within the nucleus. It is
this normal mix of Fock states of the incident particle which
leads to taking $\langle\sigma\rangle$ in the exponent.

Returning to the high energy case, we see that as the separation
$\vec r$ is frozen, each value of $r$ corresponds to a diffractive
eigenstate $|\phi_k\rangle$ (in the notation of (\ref{eq:a4}))
with cross section $\sigma_k=\sigma(\vec r)$. If there is a large
probability of inelastic scattering, that is $\sigma T > 1$, then
the difference between (\ref{eq:gammastarA}) and
(\ref{eq:lowenergyamplitude}) is significant, and leads to a large
diffraction dissociation cross section $\sigma_{SD}$ of
(\ref{eq:a7}).

In the above example we are dealing with a continuous set of
diffractive eigenstates $|\phi_k\rangle$, each characterised by a
different value of $\vec r$. Another example is to consider just
two diffractive channels~\cite{BTM}--\cite{Alberi} (say, $p,
N^*$), and assume, for simplicity, that the elastic scattering
amplitudes for these two channels are equal.  Then the $T$ matrix
has the form
\be \label{eq:a8} {\rm Im}~T \; = \; 1 \: - \: e^{- \Omega/2}, \ee
where the eikonal matrix $\Omega$ has elements
\be \label{eq:a9} \Omega_{f^\prime i^\prime}^{fi} \; = \; \Omega_0
\: \omega^{fi} \: \omega_{f^\prime i^\prime}. \ee
The individual $\omega$ matrices, which correspond to transitions
from the two incoming hadrons, each have the form
\be \label{eq:b9}
\omega \; = \; \left (
\begin{array}{cc} 1 & \gamma \\ \gamma & 1 \end{array} \right ).
\ee
The parameter $\gamma (s, b)$ determines the ratio of the
inelastic to elastic transitions.  The overall coupling $\Omega_0$
is also a function of the energy $\sqrt{s}$ and the impact
parameter $b$. It is assumed here that the diagonal elements for
both channels are equal. This leads to a simplification of the
formulas.

With the above form of $\omega$, the diffractive eigenstates are
\be \label{eq:c9} | \phi_1 \rangle \; = \; \frac{1}{\sqrt{2}}
\left (| p \rangle + | N^* \rangle \right ), \quad\quad | \phi_2
\rangle \; = \; \frac{1}{\sqrt{2}} \left (| p \rangle - | N^*
\rangle \right ). \ee In this basis, the eikonal has the diagonal
form
\be \label{eq:d9} \Omega_{m^\prime n^\prime}^{mn} \; = \; \Omega_0
\: d^{mn} \: d_{m^\prime n^\prime}, \ee
where $m, n = \phi_1,
\phi_2$ and
\be \label{eq:e9} d \; = \; \left (
\begin{array}
{cc} 1 + \gamma & 0 \\ 0 & 1 - \gamma
\end{array}
\right ). \ee
In the case where $\gamma$ is close to unity, $\gamma = 1 -
\varepsilon$, one of the eigenvalues is small. As was mentioned
above, in QCD the diagonal states are related to the colourless
$\qq$ states with definite transverse size, $r$. Small eigenvalues
correspond to the states of small size ($\sigma \sim r^2$). Thus
the $s$~channel view of diffraction is convenient to incorporate
$s$~channel unitarity. However it needs extra dynamical input to
predict the $s$ and $b$~dependences of the different diffractive
processes.

\subsection{Diffraction from a $t$ channel viewpoint}

The $t$ channel approach is based on the Regge model for
diffractive processes. In this approach diffractive processes are
mediated by the exchange of a Pomeron ($P$) -- the leading Regge
pole with vacuum quantum numbers (Fig.~1). The Pomeron plays the
role of an exchanged `particle', and gives factorizable
contributions to scattering amplitudes.

Diffractive processes (Fig.~1) are characterized by a large
rapidity gap between groups of produced particles.
\begin{figure}[h]
\begin{center}
\epsfig{figure=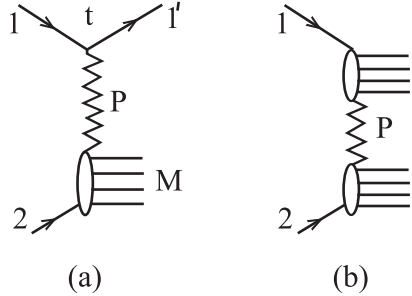,height=2in} \caption{Single and double
diffractive dissociation with a large rapidity gap represented by
Pomeron exchange.}
\end{center}\end{figure}
For example, for the (single) diffractive dissociation of
particle~2 into a system of mass $M$, the rapidity gap between
particle $1^{\prime}$ and the remaining hadrons is
\be \Delta y = \ln\left(\frac{s}{M^2}\right) =
\ln\left(\frac{1}{x_P}\right), \label{eq:Delta y} \ee
where $x_P = 1-x$ and $x=p_{1'L}/(p_{1'L})_{\rm max}$ is the
Feynman $x$ variable of particle $1'$. In general, the masses,
$M_i$, of the diffractively excited states, produced in high
energy $\sqrt s$ collisions, can be large. The only condition for
diffractive dissociation is $M_i^2\ll s$.

In the Regge pole model, the cross section for the inclusive
single diffractive dissociation process of Fig.~1(a) can be
written in the  form
\be x_P \frac{d^2\sigma }{dx_P\,dt}\ =\
\frac{(g_{11}(t))^2}{16\pi} |G_P(\Delta y,t)|^2\:\sigma_{P2}^{\rm
tot}(M^2,t) \label{eq:diffcross} \ee
where $t$ is the (square) of the 4-momentum transfer. The Green's
function (propagator) of the Pomeron is
\be G_P(\Delta y,t)={\mathcal S}\exp[(\alpha _P(t)-1)\Delta y],
\label{eq:Green}  \ee
where
\be {\mathcal S} = \frac{1 +
\exp(-i\pi\alpha_P(t))}{\sin\pi\alpha_P(t)} \ee
is the signature factor. The quantity $\sigma _{P2}^{\rm
tot}(M^2,t)$ can be considered as the Pomeron--particle~2 total
interaction cross section~\cite{Kai}. This quantity is not
directly observable. It is defined by its relation to the
diffraction production cross section, (\ref{eq:diffcross}). This
definition is useful, however, because at large $M^2$, this cross
section has the same Regge behavior as the usual total cross
sections
\be \sigma _{P2}^{\rm tot}(M^2, t)=\sum_k g^k_{22}(0)r^k_{PP}(t)
\left(\frac{M^2}{s_0}\right)^{\alpha _k(0)-1}
\label{eq:P2crosssection} \ee
where the $r^k_{PP}(t)$ is the triple-Reggeon vertex, which
describes the coupling of two Pomerons to the Reggeon $k$.

Thus, in this region where the dissociating system has a large
mass $M$, satisfying $s\gg M^2\gg m^2$, the inclusive diffractive
cross section is described by the triple-Regge diagrams of Fig.~2.
\begin{figure}[h]
\begin{center}
\epsfig{figure=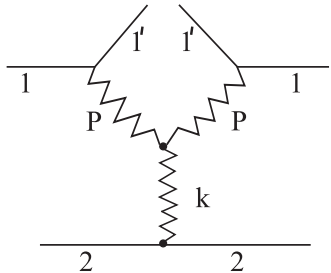,height=2in} \caption{Triple-Regge diagrams
with $k=P$ or $f$ exchange describing high energy ($M^2\gg m^2$)
Pomeron--particle~2 scattering.}
\end{center}\end{figure}
Therefore, on inserting (\ref{eq:P2crosssection}) into
(\ref{eq:diffcross}), we have
\be \frac{d^2\sigma}{dx\,dt}=\sum_k \Gamma_k(t)(1-x)^{\alpha
_k(0)-2\alpha_P(t)} \left(\frac{s}{s_0}\right)^{\alpha _k(0)-1}.
\label{eq:large_s2_diffcross} \ee

The Pomeron and $f$ couplings to the proton, and the triple-Regge
vertices $r^P_{PP}, r^f_{PP}$, have been determined from analyses
of experimental data on the diffractive production of particles in
hadronic collisions (see the review in~\cite{Kaidiff}). These
couplings specify the factors $\Gamma_k(t)$ in
(\ref{eq:large_s2_diffcross}).

In impact-parameter space, Regge amplitudes have a gaussian form
at asymptotic energies. On the other hand, for inelastic
diffraction we expect a peripheral form, which follows from
unitarity in the $s$~channel picture. Note also that, since the
Pomeron has intercept $\alpha_P(0)> 1$, the cross sections of
diffractive processes, (\ref{eq:large_s2_diffcross}), increase
with energy faster than the total cross sections, and thus lead to
a problem with unitarity. However, in Regge theory it is necessary
to take into account, not only Regge poles, but also Regge
cuts~\cite{Mandelstam,GPTM}, which correspond to the exchange of
several Regge poles in the $t$~channel. These contributions
restore the unitarity of the theory. For inelastic diffraction
they lead to a peripheral form of the impact-parameter
distributions and yield the appropriate energy dependence of the
corresponding cross sections.

How is this realized in Regge theory? The explanation is based on
a technique to evaluate Reggeon diagrams, introduced by
Gribov~\cite{Gribov}, which allows the calculation of
contributions of multi-Pomeron cuts in terms of Pomeron exchanges
in the amplitudes of diffractive processes.

\subsection{Compatibility of the $s$ and $t$ channel viewpoints:\\
Gribov's Reggeon calculus and the AGK cutting rules}

Gribov's technique uses unitarity and analyticity of
Reggeon--particle amplitudes, which follows from an analysis of
Feynman diagrams~\cite{Gribov}. For example the amplitude of
two-Pomeron exchange in the $t$~channel (Fig.~3a) can be written
as a sum over all diffractive intermediate states in the
$s$~channel (Fig.~3b).
\begin{figure}[h]
\begin{center}
\epsfig{figure=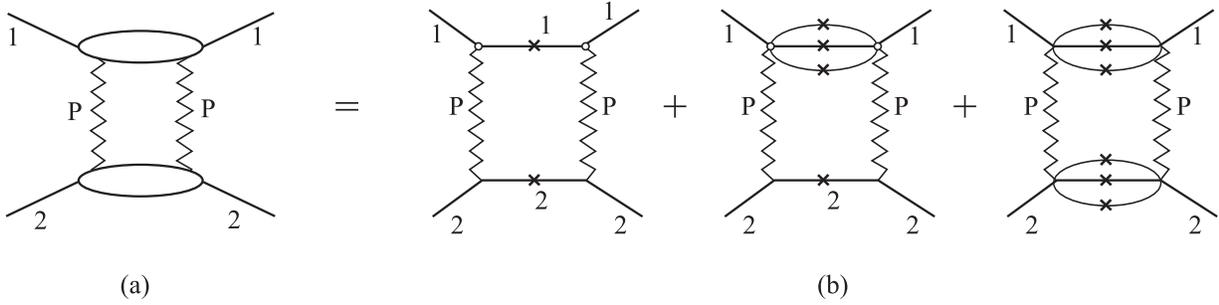,height=1.6in} \caption{Two-Pomeron
exchange in the $t$ channel expressed as the sum over all
diffractive intermediate states in the $s$ channel. The crosses
indicate that the particles are on the mass shell.}
\end{center}\end{figure}
This result can be extended to multi-Pomeron exchanges in the
$t$~channel. For the single ($s$) channel case, the summation of
all elastic rescatterings leads to the well known eikonal formula
with eikonal given by Fourier transform of the
single-Pomeron-exchange diagram. For several diffractive channels
the resulting amplitudes can be written in the matrix eikonal form
of the type shown in (\ref{eq:a8}) and (\ref{eq:a9}). In this way
a connection between the $s$~channel view of diffraction and Regge
theory is established.
Note that only for weak coupling can the Pomeron cuts be
neglected, and just the first term in the expansion be retained.
Another powerful tool for the calculation of, not only
diffractive, but inelastic processes, which contribute to total
cross section, is the AGK (Abramovsky, Gribov,
Kancheli~\cite{AGK}) cutting rules. By applying these rules, it is
possible to show the selfconsistency of the approach, which was
absent in the pure Regge pole model. Consider a diagram where
elastic scattering is mediated by the exchange of $k$ Pomerons.
The AGK cutting rules specify the coefficients $c_k^n$ arising
when $n$ of these Pomerons are cut. Recall that these Pomeron cut
discontinuities give the corresponding inelastic contributions to
$\sigma_{\rm tot}$. The terms with $n=0$ correspond to the
diffractive cutting of the diagram (that is the cut is between the
Pomeron exchanges, and not through the Pomerons themselves). These
$n=0$ terms have the coefficients
\be c_k^{n=0} = 2^{k-1} - 1.  \label{eq:coeffs} \ee
For two-Pomeron exchange, $k=2$, the coefficients are $+1$, $-4$,
$+2$ according to whether $n=0$, 1 or 2 Pomerons are cut,
respectively.

We illustrate the compatibility between the $s$~channel view of
diffraction and the $t$~channel approach with multi-Pomeron cuts,
using a simple example with two diffractive channels. In this case
the $T$~matrix has the form (\ref{eq:a8})--(\ref{eq:b9}) with
function $\Omega_0$ given by
\be \label{eq:Omega_0} \Omega_0  =   \frac{(g_{pp}^P)^2
(s/s_0)^\Delta}{4 \pi B}  e^{-b^2/4B}\ =\ \frac{(g^P_{pp})^2}{4\pi
B}\exp\left(\Delta\ln\left(\frac{s}{s_0}\right) -
\frac{b^2}{4B}\right), \ee
where $B$ is the slope of the Pomeron amplitude,
\be \label{eq:a18} B =  \textstyle{\frac{1}{2}} B_0  +
\alpha_P^{\prime}  \ln (s/s_0), \ee
$g_{pp}^P$ is the coupling of the Pomeron to the proton, $s_0 =
1~{\rm GeV}^2$ and $ \Delta\equiv \alpha_P(0)-1$. In practice, the
value of $\gamma$ is about 0.4. Thus as a first approximation we
may neglect higher powers of $\gamma$ in (\ref{eq:a8}). The
eikonal function $\Omega$ increases with energy as
$(s/s_0)^\Delta$, and at very high energies it follows, from
(\ref{eq:Omega_0}) and (\ref{eq:a18}), that $\Omega\gg 1$ for
values of the impact parameter satisfying\footnote{We have
neglected the slowly varying term $\alpha_P
'\ln(s/s_0)\ln\left[(g_{pp}^P)^2/4\pi\alpha'_P\ln(s/s_0)\right]$
in $R^2$.}
\be b^2 < R^2 = 4\alpha_P'\Delta \ln^2(s/s_0).
\label{eq:bsquaredinequality} \ee
 As a consequence, the
elastic and inelastic cross sections have strikingly different
forms in impact parameter space. The elastic cross section is
limited by the black disc bound, $\sigma_{\rm el}(b,s) = 1$, up to
$b\sim R$, after which it decreases rapidly. On the other hand,
the inelastic cross section depends on the off-diagonal elements
in
 (\ref{eq:a9}) and (\ref{eq:b9}),
\be \sigma_{12}(b,s) \simeq (\gamma\Omega_0)^2 e^{-\Omega_0}
\label{eq:sigma_12}, \ee
and so is only non-zero in the peripheral region of the
interaction, $b\sim R$. This example illustrates the general
difference in the suppression of the elastic and inelastic
channels due to multi-Pomeron exchanges, in agreement with
$s$~channel unitarity.

Let us illustrate the self consistency of the Gribov technique for
the evaluation of multi-Pomeron contributions, and the AGK cutting
rules, in the simplest single-channel example. In this case the
imaginary part of elastic amplitude in $b$ space has the eikonal
form
\be {\rm Im}\,T_{\rm el}(s,b)=
\sum_{k=1}\frac{(-1)^{k+1}}{k!}\left(\frac{\Omega_0}{2}\right)^k=
1-\exp(-\Omega_0/2). \label{eq:imaginary} \ee
By applying the AGK cutting rules \cite{AGK} to each term of the
series, it is possible to obtain for the diffractive (elastic)
cross section~\cite{TM}
\be \label{eq:a19}
 \sigma_{\rm el}(s,b) = \sum_{k=1}\frac{(-1)^k}{k!}\left(\frac{\Omega_0}{2}\right)^k2(2^{k-1}-1)
  = (1-\exp(-\Omega_0/2))^2
\ee
and for inelastic cross section
\be \label{eq:a20} \sigma_{\rm inel}(s,b) =
\sum_{n=1}\frac{(\Omega_0)^n}{n!}e^{-\Omega_0}=
(1-\exp(-\Omega_0)), \ee
where $n$ is the number of cut Pomerons (with any number of uncut
ones). It has a simple probabilistic interpretation. The second
expression for $\sigma_{\rm inel}$ represents the whole
probability, 1, minus the probability $\exp(-\Omega_0)$ to have no
inelastic interaction, whereas in the first expression each term
$\Omega_0^n/n!$ represents the probability of $n$ inelastic
interactions (where $n!$ accounts for the identity of the
interactions) multiplied by $\exp(-\Omega_0)$ which guarantees
that there are no further inelastic interactions.


>From (\ref{eq:imaginary})--(\ref{eq:a20}) we see that, in the
single channel case, we have
\be \sigma_{\rm el}(s,b)=({\rm Im}\,T_{\rm el}(s,b))^2 \ee
\be \sigma_{\rm el}(s,b)+\sigma_{\rm inel}(s,b)= \sigma_{\rm
tot}(s,b)=2{\rm Im}\,T_{\rm el}(s,b),\ee
as indeed it must be. Note that this self consistency could not be
achieved if the series in multi-Pomeron rescatterings would be cut
at some finite term. The consistency with $s$~channel unitarity
can be proven for an arbitrary multichannel case.

It is also instructive to consider the situation with two Pomeron
poles: soft $P_{\rm s}$ and hard $P_{\rm h}$. In the single
channel case, a summation of $s$~channel rescatterings leads again
to the eikonal formula \cite{Kwiec} with
\be \label{eq:21}
 \Omega(s,b)=\Omega_{\rm s}(s,b)+\Omega_{\rm h}(s,b)
\ee
where the $\Omega_i(s,b)$ are the Fourier transforms of the
$i$-Pomeron contributions. The hard Pomeron is usually related to
the cross section of minijet production~\cite{GLR,Kwiec}, which
rapidly increases with energy. It is interesting to note that the
unitarization via eikonal formula described above leads to a
rather weak influence of the hard component on the behaviour of
the total cross sections, because this component becomes important
only at energies when the soft component is large and the
amplitudes at small impact parameters are close to the black disk
limit. From the AGK cutting rules it is easy to prove that in this
model, the cross section of a hard process (with any number $k\geq
1$ of hard interactions plus any number of soft inelastic
interactions) is described by the eikonal expression with
$\Omega=\Omega_{\rm h}$. This is the so-called self absorption
theorem for processes with a given criterion~\cite{Blanken}. In
this case the criterion is the existence of at least one minijet.
>From the above expression it is clear that such processes will
dominate at very high energies, and that the mean number of
minijets will be determined by $\Omega_{\rm h}$.

\section{Diffractive eigenstates and parton configurations}

The partonic picture of hadronic fluctuations mentioned above
gives a new insight to processes of diffraction dissociation of
hadrons. This picture naturally appears in QCD, where each hadron
can be represented as a Fock state vector in terms of the quark
and gluon degrees of freedom. At very high energies, these
fluctuations have large lifetimes and have small variations during
the time of the  interaction. Thus configurations made of definite
numbers of quarks, antiquarks and gluons with definite transverse
coordinates can be considered as natural candidates for the
eigenstates $|\phi_k\rangle$ of diffraction. Partonic models of
diffraction were originally introduced in
refs.~\cite{Vanhove,Miet} and are now widely used within a QCD
framework (for examples and references see~\cite{Frankfurt}). In
QCD, colourless configurations of small transverse size have small
total interaction cross sections. Existence of such configurations
leads to the phenomenon known as `color
transparency'~\cite{BERTSCH,coltransp}, which, in turn, leads to
interesting effects in the diffractive interactions of hadrons and
photons with nuclei. Existence of the diffractive eigenstates with
substantially different interaction cross sections leads,
according to (\ref{eq:a7}), to large total cross sections of
diffractive processes at high energies, which are close to the
Pumplin bound (\ref{eq:PB}). The distribution, $P(\sigma_k)$, of
cross sections $\sigma_k$ of the diffractive eigenstates $\phi_k$,
has been determined from experimental data on diffractive
processes for pions and protons in Refs.~\cite{Strikman}. In the
simplest 2-channel model, discussed in Section 2.1, the broad
distribution $P(\sigma_k)$ is approximated by 2 states with
substantially different values of $\sigma_k$. In ref.~\cite{KKMR}
these states were related to the distributions of quarks and
gluons with substantially different values of Bjorken $x$ (see
also~\cite{Brodsky}). In Ref.~\cite {KKMR} two models were
considered. In the first model (model~A) it was supposed that the
valence quarks correspond to the small size component, while the
sea quarks and gluons make up the large size component. In the
second model (model~B) the weights of the two components varied
with $x$ in such a way that the small size component dominated at
$x\sim 1$, while the large size component was dominant at small
$x$.

Another interesting aspect of the partonic picture is the
existence of partonic fluctuations with large masses, $M\gg m_N$,
which correspond to multigluon configurations. In perturbative
QCD, these are related to the BFKL Pomeron~\cite{BFKL}. The
elastic scattering of such states leads to the diffractive
production of states of large mass, which in the Regge language
corresponds to the triple-Pomeron interaction. The average mass of
such states increases with energy, and at very high energies their
role in inelastic diffractive processes becomes very important.

It was emphasized above, that at very high energies when elastic
scattering becomes close to the black disk limit, there is a
strong influence of different diffractive eigenstates on the
properties of diffractive processes in the physical (hadronic)
basis. In particular, there is a strong reduction of the cross
section of large mass diffraction due to unitarity (multi-Pomeron
rescatterings) effects. Below we will consider these effects for
the diffractive production of large mass states, which contain a
hard subprocess: the so-called hard diffractive processes.

\section{Hard diffraction in $\pp$ ($pp$) and $pA$ collisions}

The hard diffractive dissociation of hadrons or nuclei provides
new possibilities for the investigation of diffraction dynamics.
We illustrate  this by  discussing some typical reactions below.
First, we describe a study of diffractive dijet production at the
Tevatron, and then we proceed to make predictions for typical hard
diffractive processes in proton--nuclear collisions that are
accessible at RHIC. The possible relevance to diffraction at the
LHC is mentioned.

\subsection{Diffractive dijet production in $\pp$ collisions}

Consider first the situation for dijet production in $pp$ ($\pp$)
collisions. It is important that the single Pomeron exchange
diagram of Fig.~4(a) can be calculated using QCD factorization for
the hard processes, together with the distributions of partons in
the proton and the Pomeron, denoted by $f_i^p$ and $f_i^P$
respectively.
\begin{figure}[h]
\begin{center}
\epsfig{figure=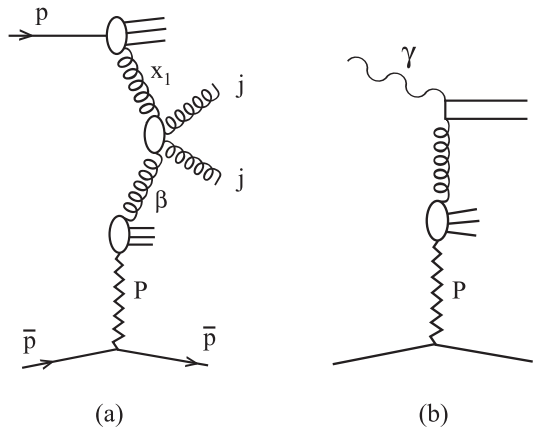,height=2.3in} \caption{(a)~The Born
diagram for diffractive dijet production at the Tevatron, and (b)~
the contribution to diffractive production at HERA driven by the
gluonic component of the Pomeron. Notice that the lower parts of
the diagrams are the same.}
\end{center}\end{figure}
The latter `diffractive' distributions can be extracted from an
analysis of hard diffraction in deep inelastic scattering
(Fig.~4(b)). Thus the cross section for diffractive dijet
production, Fig.~4(a), may be written
\be \frac{d\sigma}{dt\,dx_P} =\sum_{i,k}\int
F_P(x_P,t)f_i^P(\beta,E_T^2)f_k^p(x_1,E_T^2)\sigma_{ik}d\beta dx_1
\label{eq:c1} \ee
where $\sigma_{ik}$ is the cross section for dijet production by
partons with longitudinal momentum fractions $x_1$ and $\beta$ of
the proton and Pomeron respectively, and $E_T$ is the transverse
energy of the jets. $F_P$ is the Pomeron ``flux factor'', which,
according to (\ref{eq:diffcross}), can be written in the form
\be F_P(x_P,t) =
\frac{(g_{pp}^P(t))^2}{(16\pi)x_P^{2\alpha_P(t)-1}}. \label{eq:c2}
\ee
This approximation corresponds to the Ingelman--Schlein
conjecture~\cite{IS}. The existence of a hard scale provides the
normalization of the Pomeron term.

In the previous sections we have emphasized the importance of
multi-Pomeron contributions for diffractive particle production.
For diffractive dijet production they correspond to the diagrams
shown in Fig.~5.
\begin{figure}[h]
\begin{center}
\epsfig{figure=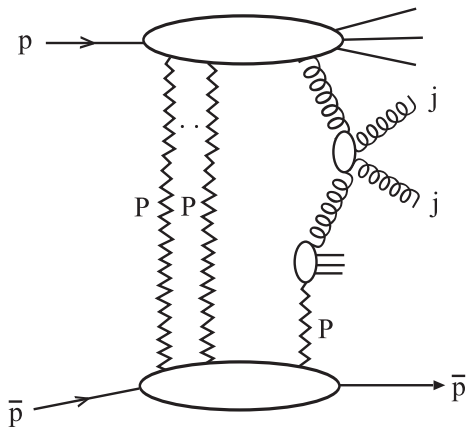,height=2.3in} \caption{Multi-Pomeron
contributions to diffractive dijet production at the Tevatron.}
\end{center}\end{figure}
These diagrams lead to a strong violation of both Regge and hard
factorization. Experimental data of CDF Collaboration at the
Tevatron~\cite{CDF} clearly indicate the breaking of hard
factorization in dijet production---the measured dijet cross
section is an order of magnitude smaller than that predicted from
(\ref{eq:c1}) using HERA data for $f_i^P$. It is even more
striking that the $\beta$ distribution for the produced dijets
increases much faster as $\beta\ra 0$, as compared with the
expected behaviour of the partonic distributions. Both
experimental observations can be reproduced in the model which
takes into account the suppression of diffractive production at
high energies due to multi-Pomeron exchanges. An essential feature
of the model, which allows the reconciliation of the observed
$\beta$ dependence with that expected from the partonic (mostly
gluonic) distributions, is the relation between the size of the
partonic fluctuation and the type (or the value of $x$) of the
active parton. The suppression factor $S_p$ in this model can be
written in the form\footnote{We use the subscript $p$ to
distinguish the suppression $S_p$ in $pp$ (or $\pp$) hard
diffractive collisions from the extra suppression $S_A$ in
$p$-nuclear collisions to be discussed in Sections~4.2 and 4.3.}
\be S_p = \frac{{\displaystyle \sum_n\int} d^2b\,
|a_{pn}|^2|{\mathcal M}_n|^2\exp(-\Omega_n(s,b))}{{\displaystyle
\sum_n\int} d^2b\,|a_{pn}|^2|{\mathcal M}_n|^2}\,, \label{eq:c3}
\ee
where $|a_{pn}|^2$ is the probability of finding the partonic
diffractive eigenstate $n \equiv |\phi_n\rangle$ in the proton
$|p\rangle$, see (\ref{eq:a4}), and $|{\mathcal M}_n|^2$ is the
probability of producing the dijet system from the eigenstate $n$.
It can depend on the type of the active parton $i$ and the value
of its momentum, $x_i$ and $\vec k_{\perp i}$.

In Ref.~\cite{KKMR} two eigenchannels were considered, $n=1,2$. It
was argued that the channel with the smaller cross section
(denoted the $S$ component) corresponds to valence quarks with
$x\sim1$, while the channel with the larger cross section is due
to sea quarks and gluons and is concentrated at smaller values of
$x$. This was denoted {\em model~A} of the diffractive
eigenstates. Of course, the model is oversimplified. There will be
part of the valence component with large size, while on the other
hand the gluons and sea quarks contribute to the small size
component. An alternative model, {\em model~B}, was introduced, in
which the partonic distributions
\be f_i(x,E_T^2)\ =\ (P_i^S + P_i^L)f_i(x,E_T^2) \ee
with $i={\rm valence,\ sea,\ glue}$; where the large and small
size projection operators have the forms
\be P_i^L = (1-x)^{n_i(Q^2)},\qquad P_i^S = 1-P_i^L, \ee
where the $n_i$ were chosen so that each component carries one
half of the whole nucleon energy and one half of the number of
valence quarks, see~\cite{KKMR}.

The functions $\Omega_n(s,b)$ in (\ref{eq:c3}) have been
parameterized in the form (see (\ref{eq:d9}))
\be \Omega_L = (1+\gamma)\Omega_0,\qquad \Omega_S =
(1-\gamma)\Omega_0, \label{eq:c4} \ee
with $\gamma = 0.4$~\cite{KMRsoft}. In practice we use a more
realistic parameterization of the $\Omega_n$, determined from the
global description of total, elastic and soft diffractive
production data~\cite{KMRsoft}. In addition to the two-channel
eikonal allowing for low mass diffraction, this analysis
incorporated pion-loop insertions in the Pomeron trajectory, and
high mass single- and double-diffractive dissociation via $s$
channel iterations of diagrams containing the triple-Pomeron
interaction.

Interestingly, the order of magnitude suppression and the $\beta$
dependence of the prediction of the CDF dijet was well reproduced
by both models~A and B for the diffractive eigenstate, see Fig.~6,
which was taken from Ref.~\cite{KKMR}.
\begin{figure}[h]
\begin{center}
\epsfig{figure=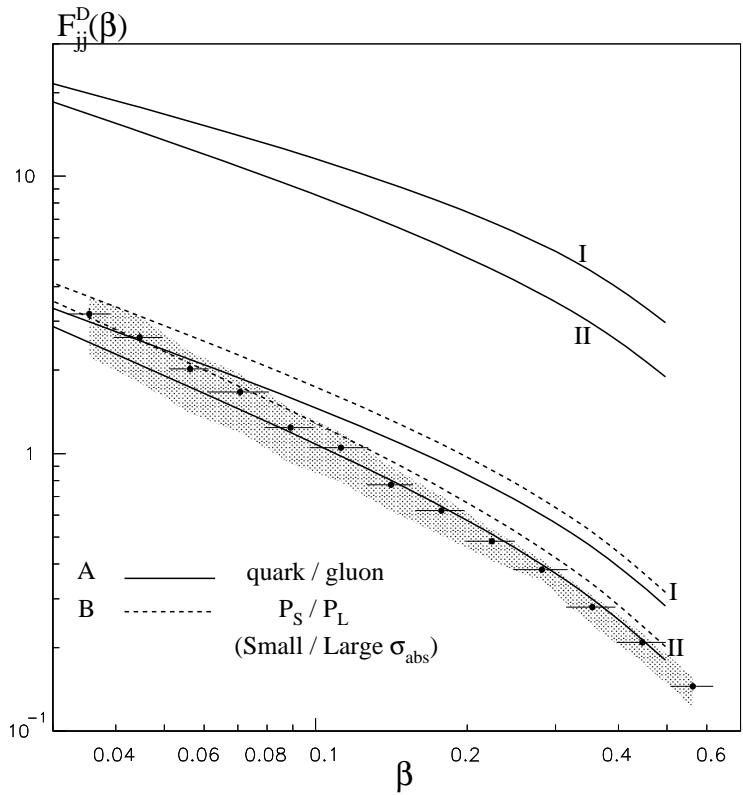,height=4.5in} \caption{The four lower
curves are the predictions for diffractive dijet production at the
Tevatron obtained from two alternative sets of `HERA' diffractive
parton distributions I~and II using models~A (continuous curves)
and B (dashed curves) to calculate the suppression factor $S_p$
of~(\ref{eq:c3}). Note that without the suppression factor the
upper two curves in the plot would be obtained. The Tevatron dijet
data (shown by the data points and shaded band) demonstrate the
importance of including the shadowing corrections. The figure is
taken from~\cite{KKMR}.}
\end{center}\end{figure}
Note that, when the new fits to H1 diffractive
data~\cite{Schilling02} are used in the approach of
Ref.~\cite{KKMR}, even better agreement with CDF dijet data is
achieved~\cite{AMST}.

A further check of this general approach has been performed in
Ref.~\cite{KKMR2}. There it was shown that the experimentally
observed~\cite{CDFjj} breakdown of factorization in  the ratio of
the yields of dijet production in single diffractive and
double-Pomeron-exchange processes is in good agreement with the
expectations of the suppression factors obtained from the
above-mentioned global analysis of diffractive
phenomena~\cite{KMRsoft}.

\subsection{Diffraction in proton--nuclear collisions}

The suppression factors, analogous to (\ref{eq:c3}), are much
stronger for diffractive $p$-nuclear processes than for $pp$ (or
$\pp$) collisions. They have a much richer structure. Their
magnitudes, their $A$ and kinematic dependences are much more
sensitive to the absorption cross sections of the proton partonic
configurations of different transverse size, see, for
example,~\cite{Fr}. Moreover, for collisions involving nuclei,
$A$, there are more possibilities of diffractive dissociation. We
will consider two types of proton--nuclear {\em hard} diffractive
processes: the incoherent and coherent production of a {\em
massive} system $h$ (accompanied by other particles, denoted by
$X$)
\be pA\ \ra\ Xh + Y_A, \label{eq:P1}\ee
\be pA\ \ra\ Xh + A \label{eq:P2},\ee
as shown in Fig.~7.
\begin{figure}[h]
\begin{center}
\epsfig{figure=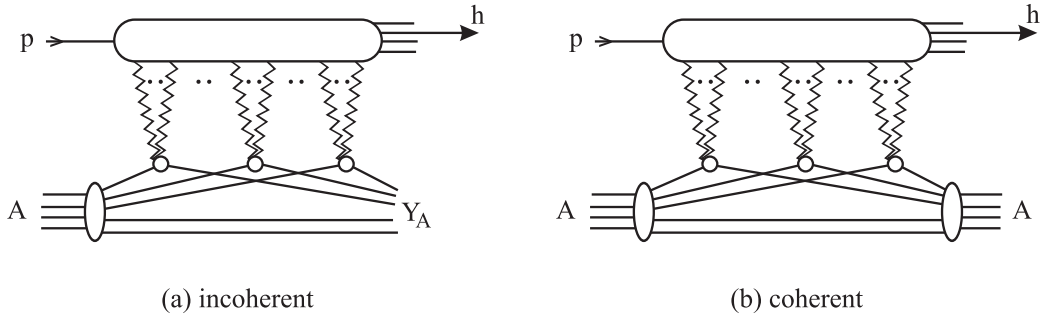,height=2in} \caption{Multi-Pomeron
exchange contributions to the incoherent and coherent diffractive
production of a massive system $h$ in high energy proton--nuclear
($pA$) collisions. In (a)~the nucleus $A$ has dissociated into a
system $Y_A$ of nucleons, whereas in (b)~it remains intact. The
dots are to indicate any number of exchanged Pomerons, and/or
interactions with any number of nucleons.}
\end{center}\end{figure}
The observable system $h$ could be a dijet, a Drell--Yan pair or a
heavy $\qq$ pair. For incoherent production, (\ref{eq:P1}), the
nucleus dissociates into a system of nucleons labelled $Y_A$. The
momentum transfer is $t\sim1/R_p^2$, where $R_p$ is the radius of
the proton. On the other hand, in process (\ref{eq:P2}) we have
the coherent dissociation of the proton, while leaving the nucleus
intact with very small momentum transfer, $t\sim1/R_A^2$, where
$R_A$ is the nuclear radius.

In the Born approximation for the interaction with a single
nucleon, the cross section, integrated over $t$, of the second
process is
\be \sigma^{\rm coh} (pA\ra Xh+A) =
\left.\frac{1}{B_A}\frac{d\sigma}{dt}\right|_{t=0} \simeq
\frac{CA^2}{A^{2/3}} = CA^{4/3}, \label{eq:sigma1D} \ee
while the cross section of the incoherent process behaves as
$\sigma(pA\ra Xh + Y_A) \sim A$. These $A$-dependences are
drastically modified by multiple rescattering (multi-Pomeron
exchanges), shown in Fig.~7.

The cross section for the incoherent hard diffraction on nuclei,
process (\ref{eq:P1}), in the Gribov--Glauber
approach~\cite{Gribov70} is described by an analogous formula to
that for $pp$ collisions,
\be \sigma(pA\ra Xh + Y_A)\ =\ \sum_n\int
d^2b\,|a_{pn}|^2\sigma^n(pp\ra h)T_A(b)\exp(-\sigma_{\rm
abs}^n(s)T_A(b)), \label{eq:P3} \ee
where, as before, the sum is over the partonic diffractive
eigenstates of the incoming proton, and $|a_{pn}|^2$ is the
probability of finding eigenstate $n\equiv|\phi_n\rangle$ in the
proton $|p\rangle$, see~(\ref{eq:a4}); $\sigma^n(pp\ra h)$ is the
corresponding cross section in proton--proton\footnote{The Pomeron
couples equally to the proton and neutron constituents of the
nucleus.} collisions for scattering through the $n^{\rm th}$
diffractive eigenstate, and $T_A(b)$ is the nuclear profile
function (\ref{eq:opticaldensity}). Here $\sigma_{\rm abs}^n(s) =
\sigma_{\rm tot}^n - \sigma_{\rm el}^n$, where the cross sections
are for proton--nucleon scattering at energy $\sqrt s$ through the
$n^{\rm th}$ eigenstate, since  elastic scattering does not change
the kinematic structure of the process. Thus the nuclear
suppression factor (sometimes called {\em nuclear transparency})
is
\be S_A^{\rm incoh}\ =\ \frac{\sigma(pA\ra Xh +
Y_A)}{A\sigma(pp\ra Xh)}\,, \label{eq:P4} \ee
where the denominator $\sigma(pp \ra Xh)$ is the corresponding
cross section for the diffractive production of the hard (massive)
system $h$ in proton--nucleon collisions. Thus $S_A^{\rm incoh}$
is the {\em extra} suppression due solely to {\em nuclear}
effects.

Coherent production on nuclei, process (\ref{eq:P2}), is possible
when the mass of the diffractively produced system, $M(hX)$,
satisfies
\be M^2/s\ \ll\ 1/m_p R_A,\label{eq:P5} \ee
where $\sqrt s$ is the proton--nucleon c.m. energy. Then the
amplitude in impact parameter space is
\be {\mathcal M}^{\rm coh} (pA\ra Xh + A)\ =\ \sum_n
a_{pn}{\mathcal M}^n(pp\ra h)T_A(b)\exp\left(-\frac{\sigma_{\rm
tot}^n}{2}T_A(b)\right), \label{eq:P6} \ee
where again the sum is over the diffractive eigenstates of the
incoming proton. Here $\sigma_{\rm abs}^n = \sigma_{\rm tot}^n$,
since even elastic scattering with $t\gtrsim 1/R_A^2$ will destroy
the nucleus. Hence the total cross section of coherent diffractive
production of $h$ is
\be \sigma^{\rm coh}(pA\ra Xh + A)\ =\ \sum_n\int d^2b\,|a_{pn}|^2
4\pi\left.\frac{d\sigma^n}{dt}\raisebox{1.6ex}{$(pp\ra
h)$}\right|_{t=0} T_A^2(b)\exp(-\sigma_{\rm tot}^n T_A(b)).
\label{eq:P7} \ee
We present below the results for the nuclear suppression factor
for the coherent diffraction production of a system $h$ in the
form
\be S_A^{\rm coh} = \left.\frac{d\sigma^{\rm
coh}}{dt}\raisebox{1.6ex}{$(pA\ra Xh + A)$}\right|_{t=0} \ \Bigg/
\ A^2\left.\frac{d\sigma}{dt}\raisebox{1.6ex}{$(pp\ra
h)$}\right|_{t=0}.\label{eq:P8} \ee
Again this is the {\em extra} suppression arising solely from {\em
nuclear} effects.

To be realistic, we present results for proton--nuclear ($pA$)
collisions in the RHIC energy regime, $\sqrt s \simeq 300$~GeV,
where $\sqrt s$ refers to the corresponding proton--nucleon c.m.
energy. We use the same model that was employed to describe hard
diffraction in $pp$ collisions~\cite{KMRsoft}. We show results
where the system $h$ is chosen to be Drell--Yan lepton pairs
(which originate from $\qq$ annihilations), and also where $h$ is
chosen to be either $\cc$ or $\bb$ pairs (which originate from
$gg$ fusion). To be specific, we use a factorization scale $\mu^2
= M^2_{\rm pair}$ where for Drell--Yan or $\cc$ production the
mass of the pair is chosen to be $M_{\rm pair} = 5$~GeV, while for
beauty production we take $M_{\bb} = 13$~GeV. The nuclear profile
function, $T_A(b)$ of (\ref{eq:opticaldensity}), was calculated
using the standard Woods--Saxon form for $\rho(r)$ with the
parameters determined from the electromagnetic form factors of
nuclei~\cite{BOHR}. In Figs.~8, 9 we show the nuclear suppression
factors (\ref{eq:P4}) and (\ref{eq:P8}) as a function of the
fraction of the longitudinal momentum of the incoming proton
carried by the system $h$ ($=l^+l^-$, $\cc$ or $\bb$).
\begin{figure}[htb]
\begin{center}
\epsfig{figure=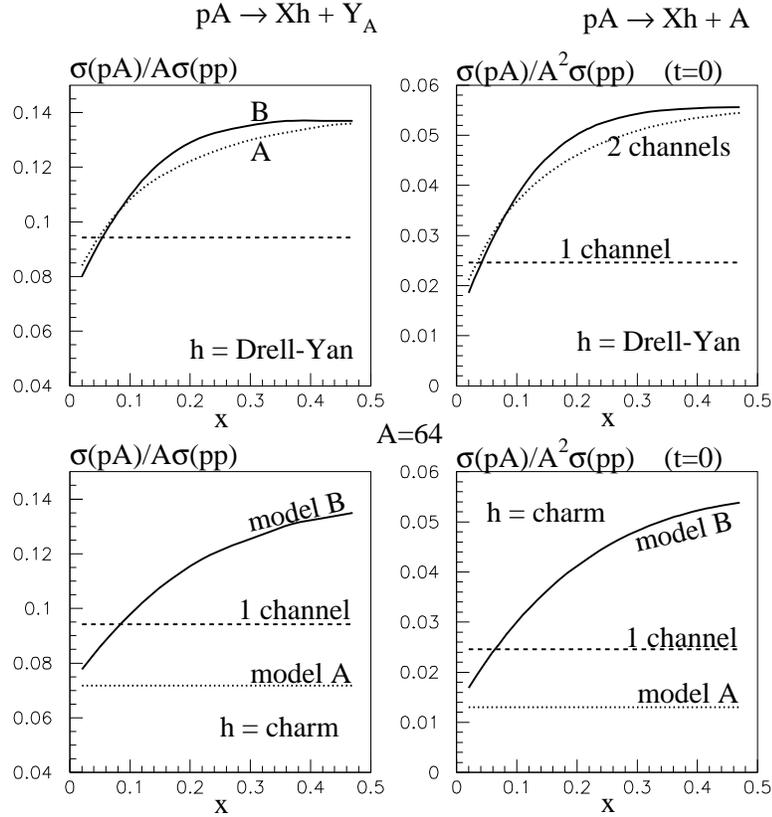,height=4.5in} \caption{The extra nuclear
suppression factors $S_A$ of (\ref{eq:P4}) and (\ref{eq:P8}) for
the incoherent ($pA\ra Xh + Y_A$) and coherent ($pA\ra Xh + A$)
diffractive production of a system $h$ carrying momentum fraction
$x$ in proton scattering on $A=64$ (Cu) nuclei at corresponding
proton--nucleon c.m. collision energy of $\sqrt s = 300$~GeV. The
dashed, dotted and continuous curves correspond to using a
single-channel and double-channel (models~A and B) eikonals
respectively.}
\end{center}\end{figure}
\begin{figure}[htb]
\begin{center}
\epsfig{figure=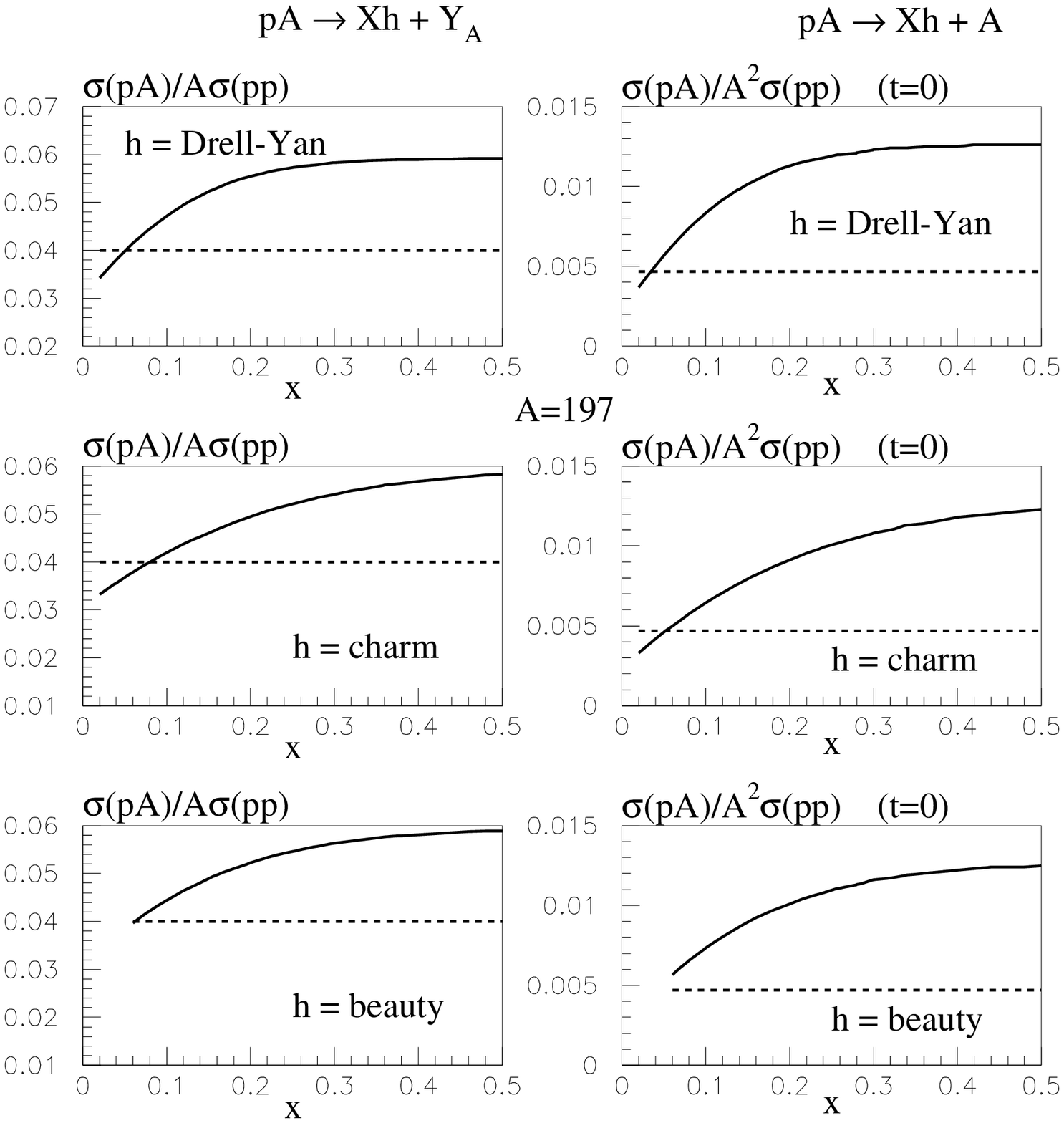,height=5in} \caption{As in Fig.~8 but for
proton collisions on $A=197$ (Au) nuclei.}
\end{center}\end{figure}
Figs.~8 and 9 correspond to nuclei with $A=64$ and $A=197$
respectively. We see that, in general, the suppression due to
nuclear effects is approximately a factor of 10 for $pA\ra Xh +
Y_A$ and a factor of 100 for $pA\ra Xh + A$. The dashed lines in
Figs.~8 and 9 correspond to the predictions of the one-channel
eikonal model, which obviously cannot represent the different
partonic configurations participating in the process. It therefore
does not depend on the specific diffractively produced system $h$
or the momentum fraction $x$ that it carries. The other curves are
the predictions of the two different two-channel eikonals
introduced in Section 4.1 --- the simple model~A, which predicts
the dotted curves, in which the valence quarks and the gluon + sea
quarks are respectively associated with the small and large size
diffractive eigenstates, $|\phi_2\rangle$ and $|\phi_1\rangle$ of
(\ref{eq:c4}); and model~B, which predicts the continuous curves,
with a more realistic partonic composition of the diffractive
eigenstates.

For models~A and B, we see that, in general, the nuclear
suppression depends on $x$. For large $x$, where the small-size,
low $\sigma_{\rm abs}$ component dominates, the suppression is
less than that expected in the pure Glauber one-channel case with
$\exp(-\langle \sigma \rangle T_A)$. On the other hand, for
smaller $x$, where more contribution comes from the large-size,
higher $\sigma_{\rm abs}$ component, there is much more
suppression. However, for charm (and beauty) production, model~A
predicts that the suppression is independent of $x$. The reason is
that the gluons have all been assigned to a single component;
moreover, as this component has $\sigma_{\rm abs} >
\langle\sigma\rangle$, the suppression is larger than the
single-channel prediction. In the more realistic model, model~B,
where the gluon is distributed between both eigenstates, the
suppression is less and the $pA$ cross section larger, and
moreover depends on $x$ --- the large $x$ gluons are concentrated
in the diffractive eigenstate with the small $\sigma_{\rm abs}$.
On the other hand, Drell--Yan production originates from valence
and sea quark annihilation and hence the difference between the
predictions of models~A and B is small.

In Fig.~9 we include the prediction for the nuclear suppression in
diffractive $\bb$, as well as $\cc$, production. The only
difference is that the scale is larger: $\mu = 13$~GeV rather than
5~GeV. By comparing $\cc$ and $\bb$ production we see the results
are not sensitive to the choice of scale. In Fig.~10 we present
the $A$~dependence for the more realistic model~B, for a large and
a small value of $x$.
\begin{figure}[h]
\begin{center}
\epsfig{figure=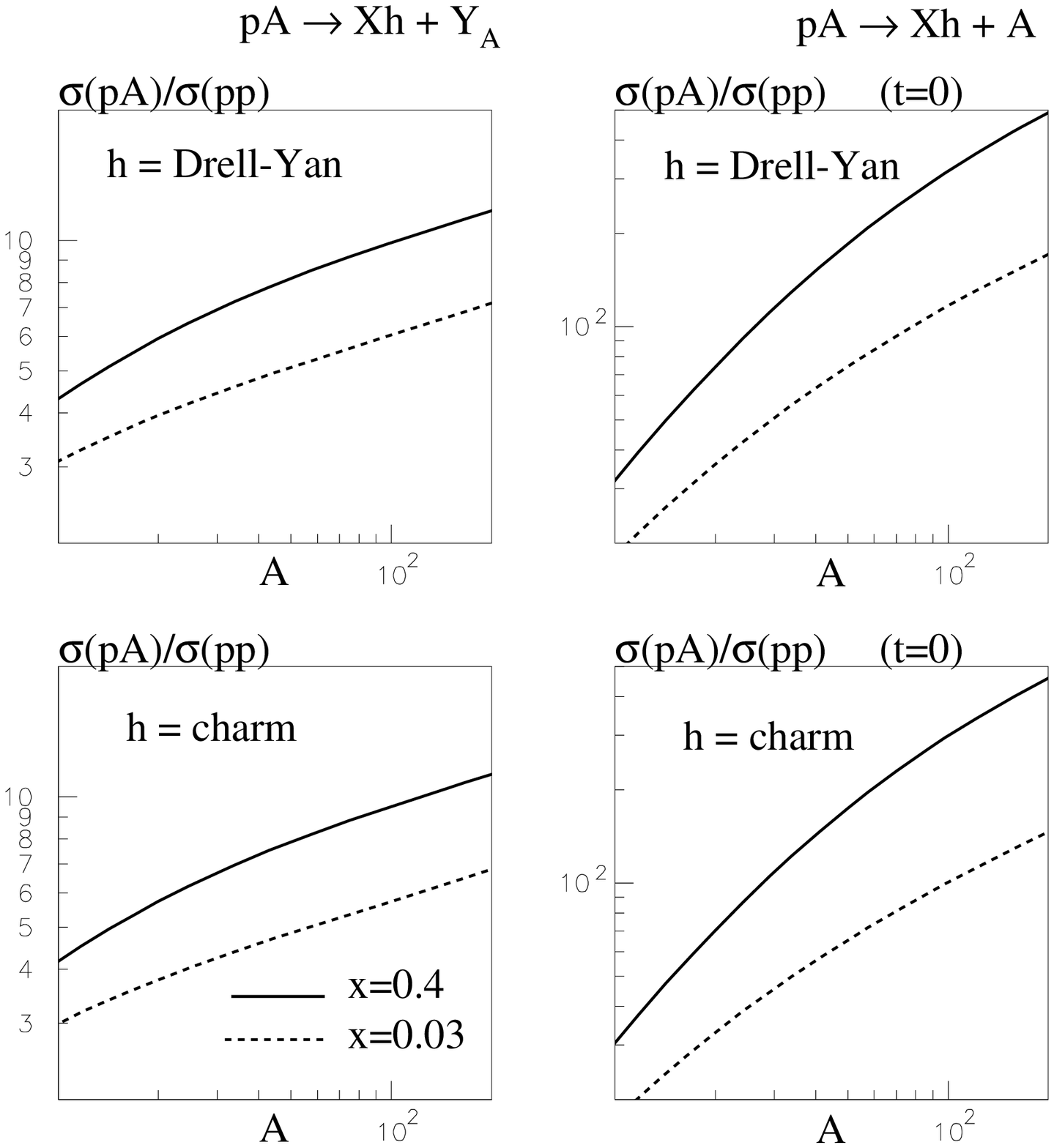,height=4.5in} \caption{The $A$ dependence
of the extra nuclear suppression factors $S_A$ of (\ref{eq:P4})
and (\ref{eq:P8}) for the diffractive production of a system $h$
in $pA$ collisions.}
\end{center}\end{figure}

\subsection{Comments on proton--nuclear diffractive processes}

The above examples demonstrate the role that diffractive
eigenstates (with different absorptive cross sections) play in the
description of diffractive processes involving nuclei. In
comparison with the over-simple single-channel Glauber approach,
the nuclear suppression factors reveal an informative rich
structure. They vary by more than a factor of two for the various
hard diffractive processes.

Figures 8--10 are qualitatively valid for all energies. The
suppression factors of nuclear origin, $S_A$, decrease only
gradually with increasing energy due to the slow growth of
$\sigma_{\rm tot}^n(s)$. Moreover, these nuclear shadowing effects
for the interactions of protons with nuclei are
similar\footnote{The similarity is not entirely complete as there
is a difference in the impact parameter distributions for $pA$ and
$pp$ collisions, and in the increase of the radius of the
interaction with energy.} to the suppression factors in
proton--proton interactions at much higher energies~\cite{KAID92}.
(Recall that for both $pA$ and $pp$ collisions we refer to the
proton--proton energy $\sqrt s$.) For instance, we see from Fig.~8
that the nuclear suppression factor in $p$-Copper collisions is
$S_A\sim0.1$ at RHIC energies of $\sqrt s = 300$~GeV, which is
comparable to the suppression $S_p$ in hard diffractive $pp$
collisions at the higher Tevatron energy, see for example, Fig.~6.
This interesting correlation may be anticipated from the
dependence of $S_A$ of (\ref{eq:P3}) and (\ref{eq:P4}) on the
exponential of the product $-\sigma^n_{\rm tot}(s)T_A(b)$. Total
cross sections increase with energy as $s^\Delta$ with $\Delta\sim
0.1$, while $T_A(b)\sim A^{1/3}$. Thus an increase of $s$ by two
orders of magnitude is equivalent to an increase in $A$ by a
factor $\sim4$. Thus, since $S_A$ for $pA$ collisions with
$A\sim64$ at RHIC energies is comparable to $S_p$ for $pp$
collisions at the Tevatron ($s\sim (3$--$4)\times10^6~\GeV^2$), we
expect that $S_p$ at the LHC ($s\sim2\times10^8~\GeV^2$) will be
close to $S_A$ for $pA$ collisions with $A\sim197$ (see Fig.~9).
So the observation of shadowing of hard diffractive processes on
heavy nuclei at RHIC may serve as a guide to the size of the
suppression factors $S_p$ which occur in similar processes in $pp$
collisions at the LHC energy. Of course, the $S_p$ factors can be
calculated directly, but confirmation from a study of
proton--nuclear scattering data at RHIC would be valuable. The
$S_p$ factors need to be known, for example, in plans to identify
New Physics phenomena in observations of such diffractive
processes at the LHC.

\section{Concluding remarks}

Investigation of both soft and hard diffractive processes provides
important information on the dynamics of strong interactions at
high energies. With increasing energy the role of unitarity
effects (which, in the Reggeon approach, materialise as
multi-Pomeron exchange interactions) becomes more and more
important. The association of eigenstates of the diffractive part
of the $T$-matrix with definite partonic configurations leads to
extra insight into the dynamics of diffractive processes in QCD.
It allows one to determine the effective sizes of the partonic
configurations in hard diffractive processes. The use of nuclear
targets provides new possibilities for detailed studies of various
aspects of both soft and hard diffractive processes. We have
demonstrated that interesting information on the properties of
partonic configurations can be obtained from a study of hard
diffractive processes in $pA$-collisions at RHIC and LHC. A
knowledge of diffraction dynamics, and especially of shadowing
effects in $pp$ and $pA$ collisions, is important for predictions
of the expected manifestations of New Physics in diffractive
processes at proton colliders, see, for example,~\cite{INC}. A
well known example is double-diffractive Higgs boson production,
see, for example,~\cite{KMR,DKMOR,VAK01}.

\section*{Acknowledgements}

It is a pleasure to acknowledge the contributions and insight that
Jan Kwieci\'nski has made to this field. We thank John Dainton,
Dave Milstead and Peter Schlein for a stimulating discussion,
Kostya Boreskov for useful information and Mike Whalley for help.
This work is supported in part by the UK Particle Physics and
Astronomy Research Council and by the grants INTAS 00-00366, NATO
PSTCLG-977275, RFBR 00-15-96786, 01-02-17095 and 01-02-17383.

 

\begin{thebibliography}{99}

\bibitem{Feinberg} E.L. Feinberg and I.Ya. Pomeranchuk, Doklady Akad. Nauk SSSR
 {\bf 93}, 439 (1953); Suppl. Nuovo Cimento v.~{\bf III}, serie~X, 652 (1956).

\bibitem{AR} M.G. Albrow and A. Rostovtsev, {\tt arXiv:hep-ph/0009336}; and references therein.

\bibitem{INC} V.A. Khoze, A.D. Martin and M.G. Ryskin, Eur. Phys. J. {\bf
C23}, 311 (2002).

\bibitem{DKMOR} A. De Roeck, V.A. Khoze, A.D. Martin, R.~Orava and
M.G.~Ryskin, Eur. Phys. J. {\bf C25}, 391 (2002); V.A.~Khoze,
A.D.~Martin and M.G.~Ryskin, Acta Phys. Polon. {\bf B33}, 3473
(2002).

\bibitem{PP1} B.E.~Cox, J.R.~Forshaw and B.~Heinemann, Phys. Lett. {\bf B540} (2002) 263.

\bibitem{BPR03} M.~Boonekamp, R.~Peschanski and C.~Royon, {\tt
arXiv:hep-ph/0301244}, and references therein.

\bibitem{Good} M.L. Good and W.D. Walker, Phys. Rev. {\bf 120}, 1857 (1960).

\bibitem{Kaidiff} A.B. Kaidalov, Phys. Rept. {\bf 50}, 157 (1979).

\bibitem{Pump82} J.~Pumplin, Physica Scripta {\bf 25}, 191 (1982).

\bibitem{DINOPRep} K.~Goulianos, Phys. Rept. {\bf 101}, 169
(1983).

\bibitem{Abram} H.~Abramowicz and A.~Caldwell, Rev. Mod. Phys.
{\bf 71}, 1275 (1999).

\bibitem{WM} M.~W\"usthoff and A.D.~Martin, J. Phys. {\bf G25},
R309 (1999).

\bibitem{Hebecker} A.~Hebecker, Phys. Rept. {\bf 331}, 1 (2000).

\bibitem{DeWolf} E.A.~De Wolf, Acta Phys. Polon. {\bf B33},
4165 (2002).

\bibitem{Pumbound} J.~Pumplin, Phys. Rev. {\bf D8}, 2899 (1973).

\bibitem{F} L.~Frankfurt, M.~Strikman and M.~Zhalov, {\tt
arXiv:hep-ph/0211336}.

\bibitem{N} I.P.~Ivanov, N.N.~Nikolaev, W.~Sch\"afer,
B.G.~Zakharov and V.R. Zoller, {\tt arXiv:hep-ph/0212176}.

\bibitem{BERTSCH} G.~Bertsch, S.J.~Brodsky, A.S. Goldhaber and J.F.~Gunion, Phys.
Rev. Lett. {\bf 47}, 297 (1981).

\bibitem{coltransp}A.B.~Zamolodchikov, B.Z.~Kopeliovich and
L.I.~Ladidus, JETP Lett. {\bf 33}, 595 (1981).

\bibitem{BTM} K.G. Boreskov, A.M. Lapidus, S.T. Sukhorukov and K.A. Ter-Martirosyan,
Nucl. Phys. {\bf B40}, 397 (1972).

\bibitem{GLM} E. Gotsman, E. Levin and U. Maor, Phys. Lett. {\bf B452}, 387 (1999);
Phys. Rev. {\bf D60}, 094011 (1999), and references therein.

\bibitem{KMR} V.A. Khoze, A.D. Martin and M.G. Ryskin, Eur. Phys.
J. {\bf C18}, 167 (2000).

\bibitem{Alberi} J.~Alberi and J.~Goggi, Phys. Rep. {\bf 74}, 1
(1981).

\bibitem{Kai} A.B.~Kaidalov, V.A.~Khoze, Yu.~F.~Pirogov and N.L.~Ter-Isaakyan, Phys. Lett. {\bf B45}, 493 (1973);
A.B. Kaidalov and K.A. Ter-Martirosyan, Nucl. Phys.
{\bf B75}, 471 (1974).

\bibitem{Mandelstam} S. Mandelstam, Nuovo Cimento {\bf 30}, 1148 (1963).

\bibitem{GPTM} V.N. Gribov, I.Ya. Pomeranchuk and K.A.
Ter-Martirosyan, Phys. Lett. {\bf 9}, 269 (1964); Vad. Fiz. {\bf
2}, 361 (1965).

\bibitem{Gribov}V.N. Gribov, ZhETF {\bf 57}, 654 (1967).

\bibitem{AGK} V.Abramovsky, V.N. Gribov and O.V. Kancheli, Sov. J.
Nucl. Phys. {\bf 18}, 308 (1974).

\bibitem{TM} K.A.~Ter-Martirosyan, Phys. Lett. {\bf B44}, 377
(1973).

\bibitem{Kwiec} A.~Capella, J.~Kwieci\'nski and J.~Tran Thanh Van,
Phys. Rev. Lett. {\bf 58}, 2015 (1987).

\bibitem{GLR} L.V.~Gribov, E.M.~Levin and M.G.~Ryskin, Phys.
Rept. {\bf 100}, 1 (1983).

\bibitem{Blanken} R. Blankenbecler et~al., Phys. Lett. {\bf B107}, 106 (1981).

\bibitem{Vanhove} L. Van Hove and K. Fialkowski, Nucl. Phys. {\bf 107}, 211 (1976).

\bibitem{Miet} H.I. Miettinen and J. Pumplin, Phys. Rev. {\bf D18}, 1696 (1978).

\bibitem{Frankfurt} K.~Agreev et~al. (FELIX), J. Phys. {\bf G28},
R117 (2002).

\bibitem{Strikman} G.~Baym, B.~Bl\"attel, L.L.~Frankfurt,
H.~Heiselberg and M.~Strikman, Phys. Rev. {\bf D47}, 2761
(1993);\\
B.~Bl\"attel, G.~Baym, L.L.~Frankfurt and M.~Strikman, Phys. Rev.
Lett. {\bf 70}, 896 (1993).

\bibitem{KKMR} A.B. Kaidalov, V.A. Khoze, A.D. Martin and M.G. Ryskin,
Eur. Phys. J. {\bf C21}, 521 (2001).

\bibitem{Brodsky} S.J.~Brodsky, Tao Huang and G.P.~Lepage, Proc. 9th
SLCA Summer Inst. on Part. Phys., 87 (1981).

\bibitem{BFKL} V.S.~Fadin, E.A.~Kuraev and L.N.~Lipatov, Sov.
Phys. JETP {\bf 44}, 443 (1976); 199 (1977);\\
I.I.~Balitsky and L.N.~Lipatov, Sov. J. Phys. {\bf 28}, 822
(1978).

\bibitem{IS}G.~Ingelman and P.E.~Schlein, Phys. Lett. {\bf B152}, 256 (1985).

\bibitem{CDF} T. Affolder et~al. (CDF Collaboration), Phys. Rev.
Lett. {\bf 84}, 5043 (2000).

\bibitem{KMRsoft}V.A.~Khoze, A.D.~Martin and M.G.~Ryskin, Eur. Phys. J. {\bf C18}, 167 (2000).

\bibitem{Schilling02} F.-P.~Schilling, {\tt arXiv:hep-ph/0210027}

\bibitem{AMST} V.A.~Khoze, A.D.~Martin and M.G.~Ryskin, 31st HEP conf., ICHEP2002, Amsterdam, {\tt
arXiv:hep-ph/0210094}.

\bibitem{KKMR2} A.B.~Kaidalov, V.A.~Khoze, A.D.~Martin and
M.G.~Ryskin, {\tt arXiv:hep-ph/0302091}, Phys. Lett. (in press).

\bibitem{CDFjj} CDF Collaboration: T.~Affolder et~al., Phys. Rev.
Lett. {\bf 85} (2000) 4215.

\bibitem{Fr} L.~Frankfurt, G.A.~Miller and M.~Strikman, Phys. Rev.
Lett. {\bf 71}, 2859 (1993).

\bibitem{Gribov70} V.N.~Gribov, Sov. Phys. JETP {\bf 29}, 483
(1969); ibid. {\bf 30}, 709 (1970).

\bibitem{BOHR} A.~Bohr and B.R.~Mottelson, Nucl. Structure, v.1,
W.A.~Benjamin Inc., New York, Amsterdam, 1969.

\bibitem{KAID92} A.B.~Kaidalov, in Proc. of XXII Int. Symposium on ``Multiparticle
Dynamics'', Santiago de Compostela, 1992, ed.~C.~Pajares (World
Scientific) p.~185.

\bibitem{VAK01} V.A. Khoze, La Thuile 2001, Results and Perspectives in Particle Physics, p.~275; {\tt arXiv:hep-ph/0105224}.
 \end{thebibliography}
 \end{document}